\documentstyle[aps,preprint,tighten,floats]{revtex}
\newcommand{\beq}    {\begin{equation}}
\newcommand{\eeq}    {\end{equation}}
\newcommand{\beqarr} {\begin{eqnarray}}
\newcommand{\eeqarr} {\end{eqnarray}}
\newcommand{\barr}   {\begin{array}}
\newcommand{\earr}   {\end{array}}
\newcommand{\no}     {\nonumber}

\newcommand{\lsim}{\mathrel{\mathop{\kern 0pt \rlap
  {\raise.2ex\hbox{$<$}}}
  \lower.9ex\hbox{\kern-.190em $\sim$}}}
\newcommand{\gsim}{\mathrel{\mathop{\kern 0pt \rlap
  {\raise.2ex\hbox{$>$}}}
  \lower.9ex\hbox{\kern-.190em $\sim$}}}

\newcommand{\mb}[1]  {\mbox{#1}}
\newcommand{\mbi}[1] {\mbox{\scriptsize #1}}

\newcommand{\br}     {\hfill\break}

\newcommand{\tb}     {\tan\beta}

\begin{document}

\draft
\preprint{
\begin{tabular}{r}
CERN--TH 95--206
\\
DFTT 47/95
\\
JHU--TIPAC 95020
\\
LNGS--95/51
\\
GEF--Th--7/95
\\
August 1995
\end{tabular}
}

\title{Neutralino dark matter in supersymmetric
models \\ with non--universal scalar mass terms.}
\author{\bf
V. Berezinsky$^{\mbox{a}}$
\footnote{
E--mail: berezinsky@lngs.infn.it, bottino@to.infn.it,
johne@cernvm.cern.ch, \\
{\phantom{$^*$E--mail: }} fornengo@jhup.pha.jhu.edu,
mignola@to.infn.it, scopel@ge.infn.it},
A. Bottino$^{\mbox{b,c}}$,
J. Ellis$^{\mbox{d}}$,
N. Fornengo$^{\mbox{e,c}}$,
G. Mignola$^{\mbox{b,c}}$
\\and
S. Scopel$^{\mbox{f,g}}$}
\address{
\begin{tabular}{c}
$^{\mbox{a}}$
INFN, Laboratori Nazionali del Gran Sasso, 67010 Assergi (AQ), Italy
\\
$^{\mbox{b}}$
Dipartimento di Fisica Teorica, Universit\`a di Torino,
Via P. Giuria 1, 10125 Torino, Italy
\\
$^{\mbox{c}}$
INFN, Sezione di Torino,
Via P. Giuria 1, 10125 Torino, Italy
\\
$^{\mbox{d}}$
Theoretical Physics Division, CERN, CH--1211 Geneva 23, Switzerland
\\
$^{\mbox{e}}$ Department of Physics and Astronomy,
The Johns Hopkins University,
\\
Baltimore, Maryland 21218, USA.
\\
$^{\mbox{f}}$
Dipartimento di Fisica, Universit\`a di Genova,
Via Dodecaneso 33, 16146 Genova, Italy
\\
$^{\mbox{g}}$
INFN, Sezione di Genova,
Via Dodecaneso 33, 16146 Genova, Italy
\end{tabular}
}
%\date{\today}
\maketitle
\begin{abstract}
Neutralino dark matter is studied in the context of  a
supergravity scheme
 where the  scalar mass terms are not
constrained by
universality conditions at the grand unification scale. We analyse in detail
the consequences of the relaxation of this universality assumption
on the supersymmetric parameter space, on the neutralino relic
abundance and on the event rate for the direct detection of relic
neutralinos.
\end{abstract}

\section{Introduction}
 The phenomenology of neutralino dark matter has been studied extensively
in the Minimal Supersymmetric extension of the Standard Model (MSSM)
\cite{Susy}.
This model incorporates the same gauge group as the Standard Model
and the supersymmetric extension of its particle content. The
Higgs sector is slightly modified as compared to that of the Standard
Model: the MSSM requires
two Higgs doublets $H_1$ and $H_2$ in order to give mass both to down-- and
up--type quarks and to cancel anomalies. After Electro--Weak Symmetry
Breaking (EWSB),
the physical Higgs fields consist of two
charged particles and three neutral ones: two scalar fields ($h$ and $H$) and
one pseudoscalar ($A$). The Higgs sector is specified at the tree level by
two independent parameters:
the mass of one of the physical Higgs fields and the ratio of the two vacuum
expectation values, usually defined as $\tan\beta=v_2/v_1 \equiv
<H_2> / <H_1>$.
The supersymmetric sector of the model introduces some other free parameters:
the mass parameters $M_1$, $M_2$ and $M_3$ for the supersymmetric partners
of gauge fields (gauginos), the Higgs--mixing parameter $\mu$ and, in general,
all the masses of the scalar partners of the fermions (sfermions).

In the MSSM it is generally assumed that
the gaugino masses  are equal at the grand unification scale $M_{GUT}$:
$M_i(M_{GUT})\equiv m_{1/2}$ and hence are related at lower scales by

\begin{equation}
M_1 : M_2 : M_3 = \alpha_1 : \alpha_2 : \alpha_3
\label{eq:GUTgaugino}
\end{equation}
\noindent
where the $\alpha_i$ (i=1,2,3) are the coupling constants of the three
Standard Model gauge groups.
The neutralinos are mass--eigenstate linear superpositions of the
two neutral gauginos ($\tilde \gamma$ and $\tilde Z$) and the two
neutral higgsinos ($\tilde H_1$ and $\tilde H_2$)
\begin{equation}
\chi = a_1 \tilde \gamma + a_2 \tilde Z + a_3 \tilde H_1 + a_4 \tilde H_2\;.
\end{equation}
\noindent
The neutralino sector depends, at the tree--level, on the following
(low--energy)
parameters: $M_1= (5/3) \tan^2 \theta_W M_2$, $M_2 \simeq 0.8\, m_{1/2}$,
$\mu$ and
$\tan\beta$. Neutralino properties are naturally discussed in the
($m_{1/2}$, $\mu$) plane, for a fixed value of $\tan\beta$. As an
example, in Fig.1 the lines of constant mass for the lightest neutralino
($m_{\chi}$)
and constant gaugino fractional weight ($P \equiv a_1^2 + a_2^2$) are plotted
in the ($m_{1/2}$, $\mu$) plane for $\tan\beta=8$.
We observe that the mass of the lightest neutralino increases
from the bottom left to the top right, while the
neutralino composition changes from higgsino dominance in the top--left
region of the plane to gaugino dominance in the bottom--right.
The regions forbidden by accelerator data are also displayed in Fig.1.

The low--energy MSSM scheme is a purely phenomenological approach, whose
basic idea is
to impose as few model--dependent restrictions as possible.
In this approach the lightest neutralino is a favourite candidate for
cold dark matter. This scheme
has been employed extensively  in the analysis of
the size and the relevance of
various possible signals of neutralino dark matter:
direct detection \cite{thdir,elflor,direct},
signals due to neutralino annihilation
in celestial bodies, namely the Earth and the Sun \cite{bodies,ap3},
and signals from neutralino
annihilation in the galactic halo \cite{halo}.
The MSSM provides a useful framework
in which neutralino phenomenology may be analysed without strong
theoretical prejudices which could, {\it a posteriori}, turn out  to be
incorrect. This scheme is also frequently
employed in analyses of the discovery potential of future
accelerators \cite{Baer}.

At a more fundamental level, it is natural to implement this
phenomenological scheme within the supergravity framework
\cite{sugra,bbo,diehl}. One
attractive feature of the ensuing model is the connection between
soft supersymmetry breaking and EWSB, which would then be induced radiatively.
The essential elements of the model are described by
a Yang--Mills Lagrangian, the
superpotential, which contains  all the Yukawa interactions
between the standard and supersymmetric fields, and by the soft--breaking
Lagrangian, which models the breaking of  supersymmetry.
Here we only recall the soft supersymmetry breaking terms
\begin{eqnarray}
&-{\cal L}_{soft}& =
\displaystyle \sum_i m_i^2 |\phi_i|^2
\no \\
&+&  \left\{\left[
A^{l}_{ab} h_{ab}^{l} \tilde L_a H_1 \tilde R_b +
A^{d}_{ab} h_{ab}^{d} \tilde Q_a H_1 \tilde D_b +
A^{u}_{ab} h_{ab}^{u} \tilde Q_a H_2 \tilde U_b +\mb{h.c.} \right] -
B \mu H_1 H_2 + \mb{h.c.}      \right\}
\no \\
&+& \displaystyle \sum_i M_i
(\lambda_i \lambda_i + \bar\lambda_i \bar\lambda_i)
\label{eq:soft}
\end{eqnarray}
\noindent
where the $\phi_i$ are  the scalar fields, the $\lambda_i$ are the
gaugino fields, $H_1$ and $H_2$ are the two Higgs fields,
$\tilde Q$ and $\tilde L$
are the doublet squark and slepton fields, respectively,
and $\tilde U$, $\tilde D$ and
$\tilde R$ denote the $SU(2)$--singlet fields for the up--squarks,
down--squarks and sleptons. In Eq.(\ref{eq:soft}), $m_i$ and $M_i$ are the mass
parameters of the scalar and gaugino fields, respectively, and $A$ and
$B$ denote trilinear and bilinear supersymmetry breaking parameters,
respectively. The
Yukawa interactions are described by the parameters $h$, which
are related to the masses of the standard fermions by the usual
expressions, {\em e.g.}, $m_t = h^t v_2$.

 The supergravity  framework  is usually  implemented with a
number of restrictive assumptions about unification at $M_{GUT}$:

     i) Unification  of the gaugino masses:
        $M_i(M_{GUT}) \equiv m_{1/2}$,

     ii) Universality of the scalar masses with a common mass denoted by
     $m_0$: $m_i(M_{GUT})$ \hfill \break
    \indent \phantom{ii)\ }  $ \equiv m_0$,

    iii) Universality of the trilinear scalar couplings:
         $A^{l}(M_{GUT}) = A^{d}(M_{GUT}) = A^{u}(M_{GUT})$ \hfill \break
    \indent \phantom{iii)\ }  $\equiv A_0 m_0$. \hfill\break

\noindent
These conditions have strong consequences for low--energy supersymmetry
phenomenology, and in particular for the properties of the neutralino
as dark matter particle.
Typically,
the lightest neutralino is constrained to regions
of gaugino dominance, that entail a large relic abundance (in wide
regions of the parameter space $\Omega_{\chi} h^2$ exceeds the
cosmological upper bound) and a small direct detection rate for
neutralino dark matter.
Indirect signals from the neutralino, such as high--energy neutrinos
from the Earth and Sun, and the products of annihilation in the halo,
are practically undetectable \cite{diehl}.

  The above assumptions, particularly ii) and iii), are not very solid,
since universality may occur at a scale higher
than $M_{GUT}$, {\em i.e.}, the Planck scale or string scale \cite{comm},
in which case renormalization above $M_{GUT}$ may weak universality in the
$m_i$, {\em e.g.}, between scalars in $\underline{\bar{5}}$ and
$\underline{10}$ representations of $SU(5)$ \cite{ehnt}. Moreover,
in many
string models the $m_i$'s are not universal even at the string scale.

  In a number of recent works \cite{OlPok,others}, deviations
from some of the
unification conditions have been considered. In particular, in
Ref.\cite{OlPok} phenomenological consequences for neutralinos of a
relaxation
of assumption ii)  have been analysed in the regime of large values of
tan $\beta$. It has been shown that deviations from condition ii) may
entail a changeover in neutralino composition from a gaugino--like state
to a higgsino--like state (or at least to a higgsino--gaugino mixed state),
with important consequences for neutralino phenomenology.

In this paper, we first explore,
over the full range of tan $\beta$,
the various scenarios which may occur when condition ii) is relaxed,
with an approach which is similar to the one adopted in the
large--$\tb$ analysis of Ref.\cite{OlPok}. We then
discuss in detail the ensuing consequences for neutralino dark
matter, with particular emphasis for its direct detection.

In the following, we  first discuss
which constraints can be applied to the parameters
when specific physical requirements are imposed. In Sect.II, we
summarize the conditions implied by radiative EWSB and define the type
of departure from universality examined in this paper.
Then, in Sect.III we establish some upper bounds on the supergravity parameters
by requiring that radiative
EWSB does not occur with excessive fine tuning among different terms.
In Sect.IV we analyse in detail the
constraints due to the requirement that EWSB takes place radiatively.
Subsequently, in Sect.V
cosmological constraints, derived from the evaluation of the neutralino
relic abundance, are discussed.  Other
constraints, from experimental data on $b \rightarrow s  \gamma $ processes
and on the mass of the bottom quark $m_b$,
are applied in Sect.VI. In Sect.VII the
effects of these various constraints
are first displayed in the ($m_{1/2}$, $m_0$) plane for fixed $\tb$ and
$A_0$, and then shown in the ($m_{1/2}$, $\mu$) plane, which provides the
most useful representation for discussing neutralino
phenomenology. We recall some specific properties of the neutral Higgs
bosons in Sect.VIII. Finally, in Sect.IX event rates for direct detection
of neutralino dark matter are discussed. Conclusions are presented in
the last Section.

\section{Radiative EWSB}
We recall that the tree--level scalar potential for the neutral Higgs fields
may be written in the form
\begin{equation}
V_0 = (M_{H_1}^2+\mu^2) |H_1|^2 + (M_{H_2}^2+\mu^2) |H_2|^2 -B\mu (H_1 H_2 +
\mb{h.c.}) + \mb{quartic D terms.}
\label{eq:higgspot}
\end{equation}
\noindent
The parameters of this potential must obey the following physical conditions:
\begin{equation}
\sin 2 \beta = \frac {-2B\mu} {M_{H_1}^2+M_{H_2}^2+2 \mu^2}
\label{eq:s2beta}
\end{equation}
\begin{equation}
M_Z^2 = 2 \frac {M_{H_1}^2-M_{H_2}^2 \tan^2 \beta}
{\tan^2 \beta -1} - 2 \mu^2
\label{eq:mz}
\end{equation}
\begin{equation}
M_A^2 = M_{H_1}^2+M_{H_2}^2+2 \mu^2 > 0~.
\label{eq:pinco}
\end{equation}
\noindent Here $M_A$ is the mass of the CP--odd neutral Higgs boson (see
Sect.VIII below), and
eq.(\ref{eq:pinco}) must in fact be strengthened to
$M_A \geq (M_A)_{lb}$, where $(M_A)_{lb}$ is
the experimental lower bound \cite{LEP}.
For instance, for $\tb \gsim 3$, $(M_A)_{lb} \simeq 45$ GeV.
Notice that the sign of $\mu$ is defined according to the convention of
reference \cite{Susy}.
We  remark that although
Eqs.(\ref{eq:higgspot}--\ref{eq:pinco}) are expressed
at the tree level,
in our actual calculations 1--loop corrections
to $V_0$ \cite{oneloop} have been included.
The $M_{H_i}$'s (as well as the sfermion and the gaugino masses and
the parameters {\it A, B} and $\mu$) evolve from
the $M_{GUT}$ scale down to the  $M_Z$ scale according to the
Renormalization Group Equations (RGE's).
This is how Eq.(\ref{eq:mz}) may be satisfied, even
if $M_{H_1}$ and $M_{H_2}$ are equal at $M_{GUT}$.

In this work we
consider deviations from universality
in the scalar masses at $M_{GUT}$, which split $M_{H_1}$ from
$M_{H_2}$. This effect is parameterized as
\begin{equation}
M^2_{H_i}(M_{GUT}) = m_0^2(1+\delta_i)~.
\label{eq:nonuniv}
\end{equation}
\noindent
The parameters $\delta_i$ which quantify the departure from universality for
the $M^2_{H_i}$ will be varied in the range ($-1$,$+1$), but are taken to be
independent of the supersymmetry parameters. This is an
{\it Ansatz}, since, when evolving the scalar masses from the unification scale
(Planck scale or string scale) to the GUT scale $M_{GUT}$, the
deviation parameters are in general functions of all the supersymmetry
parameters \cite{mark}.

Following a common procedure, Eq.(\ref{eq:s2beta}) is used to replace the
parameter  $B$ by $\tan \beta$. Thus the set of independent parameters
becomes
$m_{1/2}$, $m_0$, $A_0$, $\tan \beta$, and $\mu^2$ is given in terms of these
parameters by Eq.(\ref{eq:mz}), suitably corrected by 1--loop effects:
only the sign of $\mu$ remains
undetermined. Obviously, values of $\mu^2$ are accepted only if
they exceed the experimental
lower bound $\mu_{lb}^2$, which is derived from the lower limit
on the chargino mass \cite{LEP}: $|\mu_{lb}| \simeq 45$ GeV.

We have solved the RGE's using the 1--loop beta functions
including the whole supersymmetric particle spectrum
from the GUT scale down to $M_Z$,
neglecting the possible effects of intermediate thresholds.
Two--loop and threshold effects on the running of the gauge and Yukawa
couplings are known not to exceed 10\% of the final result \cite{lang}.
While this is
of crucial importance as far as gauge coupling unification is
concerned \cite{lang}, it is a second--order
effect on the evolution of the soft masses.
Since neutralino properties are studied
over a wide range of variation for the high scale parameters, such a degree of
refinement is not required here.

In order to specify the supersymmetry phenomenology,
boundary conditions for the gauge and Yukawa couplings have to be
specified.
Low--scale values for the gauge couplings and for the top--quark and the
tau--lepton Yukawa couplings are fixed using present experimental results.
In particular, we assign for the top mass the value $m_t=178$ GeV \cite{CDF}.
In addition, we require the unification of the bottom and tau Yukawa couplings
at the GUT scale, as would be suggested by a unifying group
that includes an $SU(5)$--like structure \cite{chano}.

The values of $M_{H_1}$ and $M_{H_2}$ at the $M_Z$ scale,
obtained from the RGE's,  may be parameterized in the following way:
\begin{equation}
M^2_{H_i} = a_i m_{1/2}^2 +b_i m_0^2 + c_i A_0^2 m_0^2+
d_i A_0 m_0 m_{1/2}~.
\label{eq:polinom}
\end{equation}
\noindent
(Notice that, in our notation, all running quantities written without
any further specification are meant to denote their values at $M_Z$.)
The coefficients
in the expression (\ref{eq:polinom}) are functions of tan $\beta$
and of the $\delta_i$'s. They  are displayed in Fig.2 (a,b) for the case
of universal scalar masses, ({\em i.e.}, $\delta_i = 0$).
 The coefficients for $M_{H_2}^2$ turn out to be very stable as
functions
of $\tb$, except for small $\tb$. More precisely, $a_2 \sim -2.5$ for $\tb
\gsim
4$ with all the other coefficients much smaller (of order 0.1). As far as
$M_{H_1}^2$ is concerned, whereas $c_1$ and $d_1$ are again very
stable
(of order 0.1), $a_1$ and $b_1$ vary rapidly as functions of $\tb$.
This property of $a_1$ and $b_1$ is due to the very fast increase of
$h^b$ for increasing $\tb$.

When a departure from $m_0$ universality is introduced, the coefficients in
Eq.(\ref{eq:polinom}), except for $a_1$ and $a_2$,
become functions of the parameters $\delta_i$:
$b_1$, $c_1$ and $d_1$ depend on $\delta_1$ and $b_2$, $c_2$ and $d_2$ on
$\delta_2$. Whereas the $b_i$'s are rapidly--increasing functions of the
$\delta_i$'s, the $c_i$'s and the $d_i$'s are rather insensitive to these
parameters.

Stringent constraints on
the parameters $m_{1/2}, m_0, A_0$ and $\tan \beta$ follow from the
request that
 the ${M^2_{H_i}}$'s, evaluated from Eq.(\ref{eq:polinom}),
satisfy Eqs.(\ref{eq:mz}--\ref{eq:pinco}). Explicitly, we require that
$\mu^2$ and  $M_A^2$, given by the expressions
\begin{eqnarray}
\mu^2 &=& \frac {1} { \tan^2 \beta -1} \{(a_1 - a_2 \tan^2 \beta) m_{1/2}^2+
(b_1 - b_2 \tan^2 \beta) m_0^2 + \nonumber \\
 & & (c_1 - c_2 \tan^2 \beta) A_0^2 m_0^2 + (d_1 -
d_2 \tan^2 \beta) A_0 m_0 m_{1/2}\}- \frac {M_Z^2} {2} \nonumber \\
&\equiv& J_1 m_{1/2}^2 + J_2 m_0^2 + J_3 A_0^2 m_0^2 + J_4 A_0 m_0
m_{1/2} - \frac {M_Z^2} {2}
\label{eq:mu}
\end{eqnarray}
\begin{eqnarray}
M_A^2 &=& (a_1 + a_2 +2 J_1) m_{1/2}^2+
(b_1 + b_2 +2 J_2 ) m_0^2 + \nonumber \\
 & & (c_1 + c_2 +2 J_3) A_0^2 m_0^2 + (d_1 +
d_2 +2 J_4) A_0 m_0 m_{1/2} - M_Z^2  \nonumber \\
&\equiv& K_1 m_{1/2}^2 + K_2 m_0^2 + K_3 A_0^2 m_0^2 +
K_4 A_0 m_0 m_{1/2} - M_Z^2
\label{eq:M_A}
\end{eqnarray}
\noindent satisfy the conditions:
$\mu^2 \geq \mu_{lb}^2$, $M_A \geq (M_A)_{lb}$ mentioned earlier.

The coefficients $J_i$ and $K_i$ in
Eqs.(\ref{eq:mu},\ref{eq:M_A}) are plotted as functions of $\tb$
in Fig.2 (c,d) for the case of $m_0$ universality.
In Fig.2c we notice that all the $J_i$'s are positive, with $J_1$ dominating
over the others: for $\tb \gsim 4$, one has
$J_1 \simeq 2.4$.
As far as the coefficients $K_i$ are concerned, we see in Fig.2d that
only two of them are sizeable: $K_1$ and $K_2$. They are
both decreasing functions of $\tb$, with $K_1 > K_2$. At very large
$\tb$ these coefficients become
very small, and $K_2$ even becomes negative (but still small in
magnitude) at $\tb \gsim 50$.

In the case of non--universality, the coefficients $J_i$ and $K_i$, except
for $J_1$ and $K_1$, become functions of the parameters $\delta_i$.
We will see in Sect.IV that many important features of the supersymmetry
parameter space depend on the signs of the two coefficients $J_2$ and
$K_2$. We show in Figs.3 and 4 how their signs depend on the
values of the $\delta_i$'s. In Fig.3 the lines $J_2 = 0$
are plotted  in the ($\delta_2$, $\delta_1$) plane for a few values of $\tb$:
for each value of $\tb$,
 $J_2$ is negative in the region above the relevant $J_2 = 0$ line and
positive below.
Similarly, in Fig.4 the $K_2 = 0$ lines are displayed in the same
($\delta_2$, $\delta_1$) plane at fixed $\tb$: $K_2$ is negative above the
$K_2 = 0$ lines, and positive below.

We now make a few comments related to Eq.(\ref{eq:M_A}), since the
value of $M_A$ plays a very crucial role in a number of important
neutralino properties. This is due to the fact that many physical
processes involving neutralinos are mediated by the neutral
Higgs bosons. Thus the
value of $M_A$ determines the size of the relevant cross sections both
through $M_A^2$--dependence in propagators and, in an implicit way,
through the couplings of the $h$ and $H$ bosons to quarks and to
the lightest neutralino $\chi$ (see Sect.VIII).
As a consequence, a small value of $M_A$ has the effect of enhancing
the magnitude of the relevant cross sections.

What values of $M_A$ do we obtain from Eq.(\ref{eq:M_A})?
Because of the properties of the coefficients
$K_i$ previously analysed, $M_A$ turns out to
be a rapidly--decreasing function of $\tb$. In Fig.5, $M_A$ is
displayed  at the representative point $m_0 = 50$ GeV, $m_{1/2} =
200$ GeV (1--loop corrections to $M_A$ have been included in the
calculation). One notices that $M_A$ is $O(M_Z)$ for
$\tb \gsim 45$. This feature provides one of the most appealing
scenarios for neutralino phenomenology.

\section{Constraints due to the absence of fine tuning}
Before we exploit fully the two constraints
$\mu^2 \geq \mu_{lb}^2$, $M_A \geq (M_A)_{lb}$
to restrict the parameter space, we apply the general criterion that
the expression (\ref{eq:mu}) is satisfied without excessive tuning among the
various terms \cite{nft,bbo}. In radiative EWSB the physical value of
$M_Z$, which sets the EW scale, may be written as
\begin{eqnarray}
M_Z^2 =
 2( J_1 m_{1/2}^2 + J_2 m_0^2 + J_3 A_0^2 m_0^2 + J_4 A_0 m_0
m_{1/2} - \mu^2)~~.
\label{eq:mz2}
\end{eqnarray}
Accidental compensation (fine tuning)
among different terms in Eq.(\ref{eq:mz2}) may occur.
We explicitly require the absence of too--strong fine tuning, {\em i.e.},
 cancellations
among exceedingly large values of the parameters $m_{1/2}$, $m_0$, $A_0$
and $\mu$. Denoting by $\eta_f$ a parameter which quantifies the
degree of fine tuning, we require \cite{nft} that
\begin{eqnarray}
\left |\frac {\Delta M^2_Z} {M^2_Z}\right| & < &
\eta_f \left |\frac {\Delta x^2_i}
{x^2_i}\right |
\label{eq:nft}
\end{eqnarray}
\noindent
where $x_i$ denotes any of the previous parameters. For instance, for
$A_0 = 0$, Eq.(\ref{eq:nft}) provides the following conditions
\begin{eqnarray}
m_{1/2}^2 < \frac {\eta_f} {2 |J_1|} M^2_Z, ~~
m^2_0 < \frac {\eta_f} {2 |J_2|} M^2_Z, ~~
\mu^2 < \frac {\eta_f} {2} M^2_Z \simeq (640\mb{ GeV})^2
\label{eq:pippo}
\end{eqnarray}
\noindent
where in the last approximate equality we have taken $\eta_f = 100$, which
means that we allow an accidental compensation at the 1\% level.
The upper bound on $m_0$ depends on $\tb$ and the $\delta_i$'s, whereas that
on $m_{1/2}$ varies only with $\tb$ (because of the nature of the {\it
Ansatz} (\ref{eq:nonuniv}): see the comment after Eq.(\ref{eq:nonuniv})).

For the sake of illustration, we give some numerical examples,
taking again $\eta_f = 100$.
For $\tb = 8$, we have, for $\delta_1 = \delta_2 = 0$,
$m_{1/2} \lsim 400$ GeV, $m_0 \lsim 1.5$ TeV. For two other pairs of
values of the $\delta_i$'s, which will be discussed later on, we
obtain $m_0 \lsim 2.4$ TeV for $\delta_1 = -0.2$, $\delta_2 = 0.4$ and
$m_0 \lsim 3.0$ TeV for $\delta_1 = -0.8$, $\delta_2 = 0.2$.
At $\tb = 53$ we have $m_{1/2} \lsim 415$ GeV and
$m_{0} \lsim (1.7-1.9)$ TeV, depending on the values for the $\delta_i$'s.
These inequalities imply for the neutralino
mass $m_\chi \lsim 170$ GeV.

In the following, when graphical representations for the parameter
space are shown, we display no--fine--tuning upper bounds
obtained from the general expression (\ref{eq:nft})
with $\eta_f = 100$. These upper bounds are denoted by dashed
lines in Figs.9--14.

\section{Constraints due to radiative EWSB}
The EWSB constraints are given by the set of
Eqs.(\ref{eq:s2beta}--\ref{eq:pinco}), or equivalently
by Eqs.(\ref{eq:mu}--\ref{eq:M_A}),
together with the conditions
$\mu^2 \geq \mu_{lb}^2$ and $M_A \geq (M_A)_{lb}$.
 From these equations  the values of
$m_0$ and $m_{1/2}$ (or $\mu$ and $m_{1/2}$) are constrained and thus some
domains in the ($m_{1/2}$, $m_0$) or ($m_{1/2}$, $\mu$) planes can be excluded.
Let us start this discussion by analyzing the condition
$M_A \geq (M_A)_{lb}$, with $M_A$ given by Eq.(\ref{eq:M_A}).
For the sake of simplicity, we put
$A_0 = 0$ for the moment. To discuss the role of $M_A \geq (M_A)_{lb}$ in
placing bounds  on $m_{1/2}$ and $m_0$, we first rewrite it explicitly as
\begin{eqnarray}
K_1 m_{1/2}^2 + K_2 m_0^2  \geq  M_Z^2 + (M_A)_{lb}^2~.
\label{eq:con1}
\end{eqnarray}
\noindent
The nature of this quadratic form in the ($m_{1/2}$, $m_0$) plane obviously
depends on the signs of the two coefficients $K_1$ and $K_2$.
As we have seen in Sect.II, it turns out that, whereas $K_1$ is always
positive, the sign of $K_2$ depends on the values of $\tb$ and of the
$\delta_i$'s.
Two different  situations may occur, depending on the sign of $K_2$.
In the case $K_2 > 0$ the region allowed by (\ref{eq:con1}) is the
one above an elliptical
branch centered in the origin of the ($m_{1/2}$, $m_0$) plane. Therefore,
both parameters $m_{1/2}$ and $m_0$ are bounded from below.
When $K_2 < 0$, the region allowed by Eq.(\ref{eq:con1}) is the one
between the
$m_{1/2}$ axis and an upward--moving hyperbolic branch. Thus, whereas
$m_{1/2}$ is still bounded from below, $m_0$ is now constrained from
above. The upper bound on $m_0$ is particularly stringent when
$K_2$ is large and negative  and $K_1$ is not large.
This occurs, for instance, at very large values of $\tb$ in the
case of $m_0$ universality.

This discussion may be extended straightforwardly  to the case
$A_0 \neq 0$. In this case the constraint $M_A \geq (M_A)_{lb}$
may be written explicitly as
\begin{eqnarray}
K_1 m_{1/2}^2 + K_2 m_0^2 + K_3 A_0^2 m_0^2 +
K_4 A_0 m_0 m_{1/2} \geq  M_Z^2 + (M_A)_{lb}^2~.
\label{eq:con2}
\end{eqnarray}
\noindent
The nature of this quadratic form depends on the sign of its
determinant. When this determinant is positive, an elliptical branch
in the ($m_{1/2}$, $m_0$) plane provides lower bounds on the two variables.
On the other hand,
a negative determinant entails an upward--moving hyperbolic branch
which places an upper bound on $m_0$. These branches are part of
conics whose axes are somewhat tilted with respect to the ($m_{1/2}$,
$m_0$) axes.

Similar implications follow from the constraint
$\mu^2 \geq \mu_{lb}^2$, which may be written explicitly as
(for $A_0 = 0$)
\begin{eqnarray}
J_1 m_{1/2}^2 + J_2 m_0^2 \geq \frac {M_Z^2} {2} + \mu_{lb}^2~.
\label{eq:con3}
\end{eqnarray}
\noindent
This quadratic form may be discussed in much the same way as the one
in Eq.(\ref{eq:con1}).
{}From the properties seen in Sect.II it turns out that the coefficient
$J_1$ is always positive,
whereas the coefficient $J_2$ is positive in the universal case,
but may be negative when deviations from $m_0$ universality are
introduced. Thus it follows that the condition $\mu^2 \geq \mu_{lb}^2$
puts lower bounds on $m_{1/2}$ and either lower or upper bounds on
$m_0$, depending on the sign of $J_2$ (due to analytic properties
identical to those discussed previously below Eq.(\ref{eq:con1})).
The condition $\mu^2 \geq \mu_{lb}^2$ sets a very stringent upper
bound on $m_0$, whenever $J_2$ is negative and large in
magnitude.
The  extension to the case $A_0 \neq 0$ may be repeated here
in a way similar to the above discussion for Eq.(\ref{eq:con2}).

Thus we have seen that two important constraints,
$\mu^2 \geq \mu_{lb}^2$ and $M_A \geq (M_A)_{lb}$,
are at work in bounding
$m_{1/2}$ and $m_0$, when EWSB is required to occur radiatively.
When $J_2$ and $K_2$ are positive, the two conditions place lower
bounds on $m_{1/2}$ and $m_0$. Similar constraints are established
by the requirements that also the sfermion masses and $m_{\chi}$
satisfy the relevant experimental bounds. These last conditions are
not explicitly discussed here, but they are taken into account in our
evaluations.

It is worth emphasizing that the most dramatic impact of the
conditions
$\mu^2 \geq \mu_{lb}^2$ and $M_A \geq (M_A)_{lb}$
over the parameter space occurs when either $J_2$ or $K_2$ (or both of
them) are negative. Under these circumstances, as we have seen above,
$\mu^2 \geq \mu_{lb}^2$ and $M_A \geq (M_A)_{lb}$
may place stringent upper limits on $m_0$, bounding the neutralino
parameter space considerably. Which of the two conditions prevails
over the other depends on the specific regions of the full  parameter
space and on the values of the $\delta_i$'s.
 In Sect.VII we will illustrate the implications of these constraints
in a few specific examples.

\section{Cosmological constraint}
Let us turn now to the evaluation of the neutralino relic abundance
$\Omega_{\chi}h^2$ and to the requirement that the lightest neutralino is not
overproduced, {\em i.e.}, $\Omega_{\chi}h^2 \leq 1$.

The neutralino relic abundance $\Omega_{\chi} h^2$ is evaluated
following the standard procedure \cite{omega,omega_poles,omega1,ap1},
according to which $\Omega_{\chi} h^2$ is essentially given by
$\Omega_{\chi} h^2 \propto <\sigma_{\mbi{ann}} v>^{-1}_{\mbi{int}}$,
where $<\sigma_{\mbi{ann}} v>_{\mbi{int}}$ is the
thermally--averaged annihilation cross section, integrated from the
freeze--out temperature to the present temperature. The standard
expansion $<\sigma_{\mbi{ann}} v> = a + b x + ...$
may be employed, with $x=T/m_\chi$,
except at s--channel resonances ($Z,A,H,h$), where a more precise
treatment has to be used for the thermal average \cite{omega_poles}.
In the evaluation of
$<\sigma_{\mbi{ann}} v>$ the full set of annihilation final
states ($f \bar{f}$ pairs, gauge--boson pairs,
Higgs--boson pairs and Higgs--gauge boson pairs), as well as the
complete set of Born diagrams are taken into account
\cite{ap1}. We recall that one of the largest contributions to the annihilation
cross section is provided by diagrams with the exchange of the pseudoscalar
Higgs boson $A$. (More relevant properties of the Higgs bosons are discussed in
Section VIII.)
We note that the constraint $\Omega_{\chi} h^2 \leq 1$ is very effective
for small and intermediate values of $\tan \beta$, but is not restrictive for
large values of $\tan \beta$.
The strong restriction in the former case comes from the large value of $M_A$
implied by small and intermediate values of $\tan \beta$ (see
Fig.5) (also the couplings of $A$ to $\chi$ and fermions are small for these
values of $\tan \beta$).

We show in Figs.6--8 a few examples where $\Omega_{\chi} h^2$ is given
as a function of $m_{\chi}$ in the form of scatter plots. These
scatter plots have been
obtained by varying the parameters $m_0$ and $m_{1/2}$ on a
equally--spaced linear grid over the ranges $10~\mb{GeV} \leq m_0
\leq 2~\mb{TeV}$,
$45~\mb{GeV} \leq m_{1/2} \leq 500~\mb{GeV}$. Furthermore, we remark
that all evaluations presented in this paper are for positive values of
$\mu$, since negative values of $\mu$ are disfavoured by the
constraints due to $m_b$ and $b \rightarrow s \gamma$ processes
(see Sect.VI). The configurations shown in Figs.6--8 satisfy the
constraints due to radiative EWSB, discussed previously.

In Fig.6 is shown the case $\tb = 8$ and $\delta_i=0$.
Here, as expected
because of the intermediate value of $\tb$, many neutralino
configurations provide $\Omega_{\chi}h^2 > 1$,
whilst only a few give $\Omega_{\chi} h^2 \leq 1$.
(Also, $M_A$ is large here because of sizeable values of $K_2$
(see Fig.2d), which helps increase $\Omega_{\chi}h^2$.)
An exception occurs when
$m_\chi \simeq M_Z/2$, since in this case the annihilation cross section is
greatly enhanced due to the Z--pole contribution.

In Fig.7 we display $\Omega_{\chi} h^2$ in a case of
non--universality ($\delta_1 = -0.2, \delta_2 = 0.4$, for definiteness).
It is easier to find $\Omega_{\chi}h^2 \leq 1$
in this case, since here the departure from
$m_0$ universality implies a changeover of the neutralino composition
from the
gaugino dominance of the previous example to  higgsino dominance
(this point will be elucidated in Sect.VII). This
implies a larger $\chi$--$\chi$ annihilation cross section and
consequently  a smaller relic abundance. Thus only
a few neutralino configurations are excluded by the
$\Omega_{\chi} h^2 \leq 1$ condition.

An example for $\Omega_{\chi} h^2$ in the case of large $\tan \beta$
and $\delta_i = 0$
is shown in
Fig.8. We see that $\Omega_{\chi} h^2 \leq 1$ imposes no constraint
since, for this very large value of $\tb$, annihilation cross sections are
very large.

\section{Constraints from $\lowercase{b} \rightarrow \lowercase{s}
 \gamma$ and $\lowercase{m_b}$}
In the evaluation of the $b \rightarrow s  \gamma$  decay rate we
have included the supersymmetric
contributions arising from the charged Higgs loops and chargino loops given
in Ref.\cite{bsgamma1}. The Higgs term always adds to the
Standard Model value and usually entails too large a value for the rate. On the
other hand, the chargino contribution gives rise to a destructive
interference for $\mu > 0$ (in our convention for the sign of $\mu$).
At large $\tan\beta$ supersymmetric contributions may
be sizeable: unless the destructive interference protects the decay rate, it
can
very easily be driven out of the present experimental bounds. In the light
of this property, the positive $\mu$ scenario appears to be the favourite one
and, as already remarked, in this paper we only show results for this case.
In comparing our predictions with observations
we have taken into account that,
as discussed in Ref.\cite{bsgamma_uncert}, large theoretical uncertainties
are present, mainly due to QCD effects. In particular, predictions depend very
strongly on the choice of the renormalization scale, leading to an inaccuracy
of
order 25\%. To account for this effect we have relaxed the experimental bounds
of Ref.\cite{bsgamma_lim} by the same amount, keeping
the renormalization scale fixed
at the representative value of 5 GeV.  Thus, our requirement is that the rate
of $b \rightarrow s \gamma$ decay falls into the range
$0.8 \times 10^{-4} \leq \mbox{BR}(b \rightarrow s \gamma) \leq 5.3 \times
10^{-4}~.$

The supersymmetric corrections to the bottom mass include
contributions from bottom--squark--gluino loops and from
top--squark--chargino loops \cite{carena}.
In the present analysis, the bottom mass is computed as a function of the other
parameters and required to be compatible with the present experimental bounds.
Theoretical uncertainties in the evaluations of $m_b$ arise both from the
running of the RGE's and from assumptions about Yukawa unification.
Since our choice
is to solve RGE's at the 1--loop level and without thresholds,
we estimate an uncertainty of the order of 10\% in our prediction for
$m_b$. In addition, a relatively small departure (see Ref.\cite{barger})
from bottom--$\tau$ Yukawa unification at the GUT scale may
significantly change the bottom mass result. To take into account such
uncertainties we have chosen to weaken the bounds on $m_b$ given in
\cite{Wright} by an amount of 10\%. Thus we require  $m_b$
to fall into the range $2.7~ \mbox{GeV} \leq m_b(M_Z) \leq 3.4~ \mbox{GeV}$.

\section{Allowed regions in neutralino parameter space}
We discuss now in a few examples how the various constraints  analysed
in the previous Sections complement each other in shaping the allowed
regions in the parameter space. We start with the ($m_{1/2}$, $m_0$)
representation, and later display our results in the
($m_{1/2}$, $\mu$) plane which provides the most useful representation for
neutralino phenomenology.

Let us first clarify a few graphical conventions adopted in our
($m_{1/2}$, $m_0$) and ($m_{1/2}$, $\mu$) plots.
Regions are left empty when at least one of the following constraints is
not satisfied: i) experimental bounds on Higgs, neutralino and sfermion masses
\cite{LEP,d0},
ii) the $\chi$ is the Lightest Supersymmetric Particle (LSP),
iii) radiative EWSB and $\mu^2 \geq \mu_{lb}^2$,
$M_A \geq (M_A)_{lb}$.
Regions forbidden by the cosmological constraint ($\Omega_{\chi} h^2 \leq
1$) are explicitly denoted by dots and those disallowed by the
$b \rightarrow s  \gamma$, $m_b$ constraints
(but not by the previous ones) are denoted by crosses (crosses are displayed
only in the ($m_{1/2}$, $m_0$) plane, but not in the
($m_{1/2}$, $\mu$) plane, to simplify these plots).
The allowed domains are denoted by squares when they satisfy
$\Omega_{\chi} h^2 > 0.01$, or by diamonds otherwise in the ($m_{1/2}$, $m_0$)
plots. They are denoted by squares in the ($m_{1/2}$, $\mu$) plots,
independently of the $\Omega_\chi h^2$ value.
To simplify the discussion, we first take $A_0 = 0$.
We comment on the $A_0 \neq 0$ case at the end of this Section.

As a first example, let us consider the representative point $\tb =8$.
For this intermediate value of $\tb$,
the cosmological constraint is expected to be very effective in view of
the arguments discussed in Sect.V. This is actually the case for
universal $m_0$, when both $K_2$ and $J_2$ are positive
(see Fig.2), so that the conditions of radiative EWSB
do not set any upper limit on $m_0$ (Fig.9a).
The empty region in the lower part of these figures is forbidden by the
experimental bound on $m_{\chi}$.
As shown in this figure, in wide regions
(denoted by dots) $\Omega_{\chi} h^2 > 1$. Thus the cosmological constraint
places a very stringent upper bound on $m_0$ for
$m_{1/2} \gsim 150$ GeV. However, for smaller values of $m_{1/2}$, an
allowed horizontal region extends up to $m_0 \simeq 2$ TeV. In fact,
along  this strip, $m_{\chi} \simeq M_Z/2$ and then
$\Omega_{\chi} h^2 \leq 1$
is satisfied (see the discussion in Sect.V).

Moving away from the universal point towards a region where $J_2$
is negative, we
expect $\mu^2 \geq \mu_{lb}^2$ to be effective in  placing a
stringent upper bound on $m_0$. This is
actually the case in the example shown in Fig.10a, which refers to the
representative point
$\delta_1 = - 0.2$, $\delta_2 = 0.4$ ($J_2 = -0.07$).  Here
it is the bound $\mu^2 \geq \mu_{lb}^2$
which provides the most stringent constraint in disallowing the
large (empty) domain on the
right side. Nevertheless,
$\Omega_{\chi} h^2 \leq 1$ is still effective in excluding an
internal region that would otherwise be allowed (see the discussion below).

Keeping $\tb = 8$, we complete our discussion by considering the representative
point
$\delta_1 = - 0.8,\delta_2 = 0.2$ shown in Fig.11a,
which gives an example where $J_2$ is very small.
The peculiarity of this example
will become clear when we discuss the relevant situation in the
($m_{1/2}$, $\mu$) plane, to which we now turn.

The shape and general properties of the physical region in the
($m_{1/2}$, $\mu$) plane are
dictated by the constraints previously derived, and they are determined
most notably by $J_2$.
It is convenient to distinguish the two cases i) $J_2 > 0$ and ii) $J_2 <0$.
For case (i)
at fixed $m_{1/2}$, $\mu$ increases for increasing $m_0$ with the
consequence that the allowed physical region extends to the right of the
$m_0=0$ line in the ($m_{1/2}$, $\mu$) plane, allowing for the neutralino
only  a gaugino--dominated region.
In the case (ii) ($J_2 <0$),  starting from the $m_0=0$ line and
increasing $m_0$ at fixed $m_{1/2}$,
one moves to the left and then one may reach regions of sizeable
higgsino--gaugino mixing or even of higgsino dominance.
Case i) applies in particular to the case of $m_0$ universality
($\delta_i = 0$) for any value of $\tb$. This is clear from Fig.3,
which shows that in the ($\delta_2$, $\delta_1$) plane the origin is below
any $J_2 = 0$ line. An example of this situation is displayed in
Fig.9b (for $\tb = 8$).

However, as we have seen in Sect.II, when the assumption of
$m_0$ universality  is relaxed, then $J_2$, which in the universal
case is positive and small,
may very easily become  negative and sizeable. In this case a changeover
in neutralino composition
from an originally gaugino--like state into a higgsino--like one occurs. This
remarkable property, discussed in Ref.\cite{OlPok} for
large $\tb$, is in fact valid over the whole range of $\tb$, if
the degree of non--universality is increased for decreasing $\tb$.
An example of case ii) ($J_2 < 0$) is shown in Fig.10b, where the
allowed region extends widely into the higgsino region.
It is instructive to compare Fig.9 with Fig.10. Looking at sections
a) of these figures, we notice that changing the values of the
$\delta_i$'s from the set $\delta_i = 0$ to the set
$\delta_1 = -0.2$, $\delta_2 = 0.4$ relaxes substantially
the cosmological constraint.  Parts b) of these figures
provide the explanations for
this feature. In fact, whereas in the former case the neutralino is mainly a
gaugino,
in the latter case $\chi$ is higgsino--like or mixed. As we already
remarked, this
implies an increase of the $\chi$--$\chi$ annihilation cross section and a
reduction of the relic abundance.
The physical region also displays an extension to the right,
in the example of Fig.11b, but
here the effect is very tiny, due to a very small $J_2$ and to
the severe upper bound on $m_0$ for $m_{1/2} \gsim 180$ GeV.
This is the first case to show a very marked ($m_{1/2}$, $\mu$) correlation.

Now we turn to the case of
large $\tb$, where new features appear. First, the
$M_A \geq (M_A)_{lb}$ condition is no longer protected by large
values of $K_1$, and may become effective  in restricting the parameter
space. Secondly,
the $m_b$ and $b \rightarrow s  \gamma$ conditions are now rather
stringent over large domains and not only occasionally relevant as in the
smaller $\tb$ cases. Thirdly, the cosmological constraint is usually
overwhelmed by the other conditions. In Figs.12a, 13a, 14a we have,
for $\tb = 53$, the following sequence of examples.
i) $\delta_1 = 0, \delta_2 = 0$ (Fig.12a): here
$K_2 < 0$, $J_2 > 0$, and since $K_2$ is negative and sizeable in
magnitude, the constraint
$M_A \geq (M_A)_{lb}$ sets an extremely stringent upper bound on $m_0$
and thus forbids the wide (empty) region on the right.
ii) $\delta_1 = 0, \delta_2 = -0.2$ (Fig.13a): here one still has
$K_2 < 0$, $J_2 > 0$, but $|K_2|$ is smaller than in the previous case, so the
constraint $M_A \geq (M_A)_{lb}$ is still very effective but less compelling
than in the case (i). Also, the role of the $m_b$ and the
$b \rightarrow s \gamma$
conditions is more significant here.
iii) $\delta_1 = 0.7, \delta_2 = 0.4$ (Fig.14a): here
$K_2 > 0$, $J_2 < 0$,
$M_A \geq (M_A)_{lb}$ gives a lower
bound on $m_{1/2}$ and the $\mu^2 \geq \mu_{lb}^2$ condition
provides the  frontier of the empty domain on the right.

The ($m_{1/2}$, $\mu$) representations for large $\tan \beta$ and
for the
representative $\delta_i$ points discussed
above are displayed in Figs.12b--14b.
We start from the universal case of Fig.12b. Here we expect gaugino--dominated
configurations. However, because the values of $m_0$ are strongly limited from
above (see Fig.12a), we have the extremely correlated states shown in Fig.12b.
In the
case of Fig.13b one has $J_2>0$, and gaugino--dominated states occur. No strong
($m_{1/2}$, $\mu$) correlation shows up in this case.
The opposite case, $J_2 < 0$, is shown in Fig.14b, where
higgsino--dominated configurations appear.

It is worth adding a few comments about the
examples of Figs.11 and 12, where
the physical regions in the ($m_{1/2}$, $\mu$) plane show a very
pronounced correlation in the two variables. This feature occurs whenever
 $|J_2| m_0^2 \ll J_1 m_{1/2}^2$, {\em i.e.}, whenever $m_0$ is severely
bounded
from above and/or $|J_2|$ is very close to zero. As far as the values of
$|J_2|/J_1$ are concerned, we notice that in the universal case (see
Fig.2c), except for small values of $\tb$, $J_2/J_1 \simeq 0.04$
(in fact, for $\tb \gsim 4$,
$J_1 \simeq - a_2 \simeq 2.5$, $J_2 \simeq - b_2 \simeq 0.1$). Thus for
$\delta_i = 0$ a strong ($m_{1/2}$, $\mu$) correlation occurs whenever
$m_0 \lsim O(m_{1/2})$. This happens in the example of Fig.12,
where $m_0$ is severely bounded by the $M_A \geq (M_A)_{lb}$ condition,
and  in the case of Fig.11, where the correlation is enforced by a very
small value of $J_2$: $J_2 = 0.06$. A ($m_{1/2}$, $\mu$) correlation is also
exhibited in Fig.9b for the range $m_{1/2} \gsim 150$ GeV, where $m_0$
is bounded by the cosmological constraint.

In general, we do not consider these physical regions with a strong
($m_{1/2}$, $\mu$) correlation as unnatural, since they are usually
realized without much tuning. We recall that the size of the
coefficients $J_1$ and $J_2$ is dictated by the RGE's
with their intrinsic cancellations, and that one naturally has
$J_1= O$(a few), $J_2= O(0.1-0.01)$. As we have seen, these properties,
combined with severe upper bounds on $m_0$, are sufficient to
generate the ($m_{1/2}$, $\mu$) correlation.

We turn now to the $A_0 \neq 0$ case. First we recall that $A_0$ is constrained
in the range $|A_0| \lsim 3$ from the absence of charge and color breaking
\cite{a_0}. Thus, allowing $A_0 \neq 0$ does not change essentially the general
picture  previously discussed. The previous scenarios still occur, but at
different points in the parameter space. Two specific comments are in order
here: i) independently of its sign, $A_0$ disfavours the changeover from
gaugino dominance to higgsino dominance in the neutralino composition, ii) a
negative $A_0$ reduces the value of $M_A$ as compared to the $A_0=0$ case, and
so either provides a light $A$ boson (and hence interesting phenomenology) or
enforces a more stringent constraint on the parameter space.

\medskip
\section{Neutral Higgs Bosons}
Neutralino direct detection, to be discussed in the next Section,
is based on neutralino--nucleus scattering. In this process,
exchanges of neutral Higgs bosons play a dominant role, provided the
Higgs masses are not too heavy.
It is convenient to recall here some relevant properties
of the couplings of $\chi$ with matter via Higgs exchange. As was
already mentioned in the Introduction,
the two Higgs isodoublets $H_1$, $H_2$ yield 3 neutral Higgs
mass eigenstates:  one $\mb{CP}$--odd ($A$)
state, whose mass $M_A$ is given by
expression (\ref{eq:M_A}) and two $\mb{CP}$--even
states (of masses $M_h$, $M_H$,
$M_h < M_H$), which are obtained from
$H_1^0$, $H_2^0$ by a rotation through an angle $\alpha$
\beqarr
H &=& \cos\alpha\,H_1^0 + \sin\alpha\,H_2^0  \nonumber \\
h &=& -\sin\alpha\,H_1^0 + \cos\alpha\,H_2^0 ~.
\eeqarr
\noindent
It is important to notice here that $\alpha$ depends very
sensitively on $M_A$, being very close to
zero for $\tb \gsim 4$ and rising very fast to $\pi/2$
for $M_A \lsim  O(M_Z)$ (see Fig.15).

The angle $\alpha$ plays a crucial role in determining the size of the
neutral $h,H$--quark couplings. Here, as
we are interested in $\chi$--nucleus scattering,
we discuss explicitly only the couplings
involving the CP--even states, since $h,H$ are dominant compared to $A$.
The low--energy neutralino--quark
effective Lagrangian generated by Higgs exchange may be written
as  follows \cite{[4]}
\beq
{\cal L}_{\mbi{eff}} = \sqrt2 G_F {m_Z \over m_{h,H}^2} F_{h,H}
\sum_q k_q m_q \bar{\psi}_{\chi} \psi_{\chi} \bar{q} q~.
\label{eq:el}
\eeq
\noindent
Here $F_{h,H}$ is the ratio of the Higgs--neutralino coupling to the $SU(2)$
gauge coupling, which depends on the composition of $\chi$
\beqarr
           F_h &=& a_2 (a_3 \sin \alpha + a_4 \cos \alpha) \nonumber \\
           F_H &=& a_2 (a_3 \cos \alpha - a_4 \sin \alpha)
\eeqarr
\noindent and the $k_q$ are given, for the up--type quarks and the
down--type quarks respectively, by
\begin{eqnarray}
 \,\qquad &\qquad  H \qquad  & \qquad \qquad \quad  h \qquad \nonumber \\
k_u  \qquad &  \qquad  \sin\alpha / \sin\beta \qquad  &  \qquad
\phantom{-} \cos\alpha / \sin\beta \qquad \nonumber \\
k_d  \qquad & \qquad  \cos\alpha / \cos\beta \qquad & \qquad
- \sin\alpha / \cos\beta~.
\end{eqnarray}
\noindent
Note that, in general, since $\tb > 1$, the strength of the
coupling to the down--type quarks is bigger than the one to
the up--type quarks, and
${\cal L}_{\mbi{eff}}$ usually gets a sizeable contribution when
 the $h$ boson is exchanged
 ($h$ is lighter than $H$ and is therefore favored
because of the propagator denominator in Eq.(\ref{eq:el})) and when
$\alpha \simeq \pi/2$, {\em i.e.}, when $M_A \lsim O(M_Z)$.
When this regime does not apply, the size of ${\cal L}_{\mbi{eff}}$ is
much suppressed.

The cross section for elastic neutralino--nucleus scattering
which follows
from the effective Lagrangian (\ref{eq:el}) will be given in Sect.IX.B.

\section{Direct detection}
Much experimental activity is under way in the direct search for
neutralino dark matter and the perspectives for
significant improvements in experimental sensitivities are
encouraging \cite{mosca}.  In this class of experiments, a
relic neutralino would be detected by the amount of energy released by
its elastic scattering off nuclei in an appropriate apparatus. A
signature would be provided by a yearly modulation of the signal,
whose observations would require high statistics and
extremely good stability in the
detector response.  Here we evaluate the event rates for
this process extending previous analyses to the non--universal $\delta_i \neq
0$ case.
Various materials are being used in the current experiments and others
are under investigation for future detectors. In this paper
we analyse two of
the most interesting materials: Ge (in its natural composition)
\cite{[15],[14],klapdor} and
$^{129}$Xe \cite{[17]}.

\subsection{Differential rates}
  The nuclear recoil spectrum may be evaluated from the expression
\beq
\frac {dR}{dE_R}=
\sum_i \frac {R_{0,i}} {<E_R^{max}>} F_i^2(E_R)I(E_R)
\label{eq:dir1}
\eeq
\noindent
 where
\beq
R_{0,i}=N_T \frac {\rho_{\chi}} {m_\chi} \sigma_i <v>~.
\label{eq:dir2}
\eeq
\noindent
 In Eqs.(\ref{eq:dir1})--(\ref{eq:dir2}) we use the following
notations:
the subscript $i$ refers to the two cases of coherent and spin--dependent
effective interactions,
$N_T$ is the number of the target nuclei per unit of mass,
$\rho_\chi$ is the local neutralino matter density, and
$E_R$ is the nuclear recoil energy given by
$E_R={{m_{red}^2}}v^2(1-\cos \theta^*)/{m_N}$,
where $\theta^*$ is the scattering
angle in the neutralino--nucleus center--of--mass frame,
$m_N$ is the nuclear mass, $m_{\rm red}$ is the neutralino--nucleus
reduced mass and $v$ is the relative velocity. The maximum value of $E_R$ is
$E_R^{max}={{2m_{red}^2}}v^2/{m_N}$. Returning to
(\ref{eq:dir1}--\ref{eq:dir2}),
$F(E_R)$ denotes the nuclear form factor, and
$\sigma_i$ is the (coherent/spin--dependent) neutralino--nucleus cross section.
The factor $I(E_R)$ is given by
\beq
I(E_R)=\frac {<v^2>} {<v>}
\int_{v_{min}(E_R)}^{v_{max}}  dv \frac {f(v)} {v}
\label{eq:dir3}
\eeq
\noindent
where $f(v)$ is the velocity distribution of neutralinos in the
Galaxy, as measured in the Earth's rest frame, and
$v_{\rm min}(E_R)$ is given by
$v_{\rm min}(E_R)=({{m_N E_R}/({2m_{\rm red}^2})})^{1/2}$.
The averages
appearing in Eqs.(\ref{eq:dir1})--(\ref{eq:dir3})
denote averages over the velocity distribution in
the Earth's rest frame. An explicit formula for $I(E_R)$ in the case of a
Maxwellian velocity distribution may be found in Ref.\cite{direct}.

The differential rates to be discussed below will be
expressed in terms of the electron--equivalent energy $E_{ee}$ rather than
in terms of $E_R$. These two variables are
proportional: $E_{ee}=Q E_R$ where $Q$ is called the quenching factor: typical
values of $Q$ will be discussed shortly.

\subsection{Neutralino--nucleus elastic cross sections}

The total cross sections for neutralino--nucleus elastic scattering
have been evaluated following  standard procedures
\cite{elflor,direct,[4],[5],goodman}.
Here we only summarize some of the main properties. Neutralino--quark
scattering is described by amplitudes with
Higgs--boson exchanges and $Z$--boson exchange
in the t--channel,
and by amplitudes with
squark exchanges in the s-- and u--channels.
The neutral Higgs bosons considered here are the two
CP--even bosons: $h,H$ and the CP--odd
one: $A$, whose couplings were previously discussed in Sect.VIII.

 The relevant properties for these amplitudes are: 1) Higgs--boson
exchanges contribute a coherent cross section which vanishes only
when there is no zino--higgsino mixture in the neutralino composition
\cite{[4]},
2) $Z$--boson exchange
provides a spin--dependent cross section which receives
contributions only from the higgsino components of $\chi$, 3) squark
exchanges contribute a coherent cross section (due to zino--higgsino
mixing) as well as a spin--dependent cross section (due mainly to the
gaugino components of $\chi$)\cite{[5]}.
As examples we recall here only the expressions for the coherent
cross section due to the exchange of a Higgs boson ($h$ or $H$) and the
spin--dependent one due to Z exchange.

The former cross section is easily evaluated from the effective
Lagrangian of Eq.(\ref{eq:el}) \cite{[4]}
\beq
\sigma_{\mbi{CH}} = \frac {8 G_F^2} {\pi} \frac {m_Z^2} {m_{h,H}^4}
\alpha^2_{h,H} m_{red}^2 A^2
\eeq
\noindent
where $A$ is the
nuclear mass number and $\alpha_{h,H}$ is given by
\beq
\alpha_{h,H} = F_{h,H} I~~~~,~~~~
I=\sum_q k_q m_q \langle N|\bar{q} q |N \rangle .
\eeq
\noindent
The quantity $I$ may be expressed conveniently in terms of the
$\pi N$ sigma--term $\sigma_{\pi N}$ and of a parameter
$a$ which is related to the strange--quark content of the nucleon
$y$ by
\beq
a = y (m_s/(m_u + m_d))~~~,~~~~~ y = 2 \frac {<N|\bar{s} s|N>}
{<N|\bar{u} u+\bar{d} d|N>}~.
\eeq
One has
\beq
 I \simeq k_u g_u+k_d g_d
\eeq
\noindent
where
\beqarr
g_u &=& {4 \over {27}} \left( m_N  + {{19}\over{8}} \sigma_{\pi N} -
a \sigma_{\pi N} \right)
\nonumber \\
g_d &=& {2 \over {27}} \left( m_N  + {{23}\over{4}} \sigma_{\pi N} +
\frac {25} {2} a \sigma_{\pi N} \right) ~.
\eeqarr
\noindent
Unfortunately, the values of both the quantities $y$ and $\sigma_{\pi N}$
are somewhat uncertain. Here, for $y$ we use the central value of the
most recent evaluation: $y = 0.33 \pm 0.09$, obtained from a lattice
calculation \cite{liu}. For $\sigma_{\pi N}$, which is derived by
phase--shift analysis and dispersion relation techniques from
low--energy pion--nucleon scattering cross--sections \cite{chen,gass}, we
employ
the value of Ref.\cite{gass}: $\sigma_{\pi N} = 45$ MeV. We then find the
results: $g_u =123$ MeV, $g_d =288$ MeV (we use $2 (m_s/(m_u + m_d)) = 29$
\cite{bj}).   We note that these values further
reinforce the role of the down--type quarks as compared to the up--type
ones.

We point out that the Higgs--nucleon couplings for nucleons
bound in a nucleus may be renormalized by the nuclear
medium. As a consequence, the strength of $I$ might in principle be
reduced to some extent \cite{brown}. However, this effect is neglected here.

Now let us turn to the spin--dependent cross section due to
Z exchange. This may be cast into the usual form
\cite{[5],goodman}
\beq
\sigma_{\mbi{SD}} = \frac {8 G_F^2} {\pi} (a_3^2-a_4^2)^2 m_{red}^2
(\sum_q T_{3L,q} \Delta q)^2 \lambda^2 J(J+1)~.
\label{eq:sd}
\eeq
\noindent
In this paper we use this formula for $^{73}$Ge (this isotope is present
at the level of 7.8 \% in the natural composition of Ge) and to $^{129}$Xe. For
these nuclei we employ the values of $\lambda^2$ obtained in the
odd--group model \cite{elflor}, where only the odd nuclear species in
odd--even nuclei are explicitly taken into account. The $\Delta q$'s in
Eq.(\ref{eq:sd}) denote the fractions of the nucleon spin carried
by the quarks $q$ in the
nucleon of the odd species, and the $T_{3L,q}$'s stand for the third
components of the quark weak isospin. The values for the $\Delta q$'s
are taken from Ref.\cite{ellkar}.

   It is worth noticing that the
event rates for neutralino direct detection with the materials considered here
are largely dominated by coherent
effects in most regions of the parameter space. In the small domains
where spin--dependent effects dominate over the coherent ones the total
rates are usually too small to allow detection.
The experimental strategy of employing materials enriched in
heavy isotopes of high spin is interesting for a search for
hypothetical dark matter particles which interact with matter via
substantial spin--dependent interactions. However, this
approach does not appear to be very fruitful for neutralinos.

One more ingredient which enters the event rate in Eq.(\ref{eq:dir1})
is the nuclear form factor, which
depends sensitively on the nature of the effective
interaction involved in the neutralino--nucleus scattering.
For the coherent case, we simply employ the standard
parameterization \cite{[12]}
\beq
F(E_R)=3 \frac {j_1(qr_0)}  {qr_0} e^{-\frac {1} {2}} s^2 q^2
\label{eq:ff}
\eeq
\noindent
where $q^2 \equiv\mid{\vec {q}}\mid^2=2m_NE_R$ is the squared
three--momentum transfer, $s \simeq 1~ \rm fm$ is the
thickness parameter for
the nucleus surface, $r_0 = (r^2-5s^2)^{1/2}$, $r=1.2~A^{1/3}$ fm and
$j_1(qr_0)$ is the spherical Bessel function of index 1.

The form factor in Eq.(\ref{eq:ff}) introduces a substantial suppression in the
recoil spectrum unless $q r_0 \ll 1$. A noticeable reduction in $dR/dE_R$ may
already occur at threshold $E_R=E_R^{\rm th}=E_{\mb{ee}}^{\rm th}/Q$ when
$r_0 \sqrt{2 m_N E_R^{\rm th}}$ is not small compared to unity. The actual
occurrence of this feature depends on parameters of the detector
material: nuclear radius, quenching factor, threshold energy
$E_{\mb{ee}}^{\rm th}$. The values
of these parameters for the nuclei considered in this
paper are reported in Table I  \cite{mosca,[14],klapdor,[17]}, and
the values of $F^2(E_R^{\mb{th}})$ calculated from
Eq.(\ref{eq:ff}) are given in this same Table.
Since we consider in this paper mainly the value of the differential rate near
threshold, $F^2(E_R^{\mb{th}})$ is the most relevant quantity.
We see from the values in the Table that the reduction
introduced by the form factor is moderate
in Ge, but quite substantial in $^{129}$Xe.

In general, for the spin--dependent case there are no
analytic expressions for the form factors.
However, numerical analyses have been performed for a number of nuclei.
The general feature is that these form factors
have a much milder dependence on $E_R$ as compared to the coherent ones,
because only a few nucleons participate in the neutralino--nucleus
scattering in this case.
In our evaluations we use the results of Refs.\cite{[12],[19]} for
$^{131}Xe$ and $^{73}Ge$ respectively.

\bigskip

%Table
%\begin{center}
\begin{table}
\centering
\caption{
Characteristics of some current experiments.
In the second column is reported the quenching
factor $Q$, in the third column the electron--equivalent energy at
threshold, in the fourth the square of the form factor at threshold, and in
the last column the present experimental sensitivity.
}

\begin{tabular}{|r|r|r|r|r|}   \hline
{\em \mb{Nucleus}} & $Q$ & $E_{ee}^{th}(\mb{KeV})$ & $F^{2}(E_{R}^{th})$ &
${\mb{evts/(Kg~d~ KeV)}}
$\\  \hline
Ge\cite{[14]} & 0.25 & 2 & 0.87 & 3.0 \\ \hline
Ge\cite{klapdor} & 0.25 & 12 & 0.41 & 0.2 \\ \hline
Xe\cite{[17]} & 0.80 & 40 & 0.07 & 0.8 \\ \hline
\end{tabular}
\end{table}
%\bigskip
%\end{center}

\subsection{Local Neutralino Density}
We denote the local halo density by $\rho_l$, for which we use the estimate
$\rho_l = 0.5$ GeV cm$^{-3}$ \cite{turner}.
 For the value of the local neutralino density $\rho_\chi$ to be
used in the rate of Eq.(\ref{eq:dir2}), for each point of the model parameter
space we take into account the relevant value of the cosmological neutralino
relic density. When $\Omega_\chi h^2$ is larger than a minimal
$(\Omega h^2)_{\rm min}$  required by observational data and by large
scale structure calculations
we simply put $\rho_\chi=\rho_l$.
When $\Omega_\chi h^2$ turns out  less than
$(\Omega h^2)_{\rm min}$, the neutralino may only provide a fractional
contribution
${\Omega_\chi h^2 / (\Omega h^2)_{\rm min}}\equiv  \xi$
 to $\Omega h^2$; in this case we take
$\rho_\chi = \rho_l \xi$.
The value to be assigned to $(\Omega h^2)_{\rm min}$ is
somewhat arbitrary. Here we set it equal to 0.1.

It is worth remarking here that, due to this scaling procedure, for the
direct detection rate one has: i) $R_{0,i} \propto \rho_l \sigma_i$
for ${\Omega_\chi h^2 \geq (\Omega h^2)_{\rm min}}$ and
ii) $R_{0,i} \propto \rho_l\xi\sigma_i \propto \rho_l\sigma_i/
<\sigma_{\mbi{ann}} v>_{\mbi{int}}$
for ${\Omega_\chi h^2 < (\Omega h^2)_{\rm min}}$.
Thus the rate $R_{0,i}$ is large in the regions of the parameter space
where $\sigma_i$ is large. This is trivial in case i), but it is also
true in case ii), since when $\sigma_i$ is large also $\sigma_{\mbi{ann}}$
increases but in such a way that usually the ratio
$\sigma_i/\sigma_{\mbi{ann}}$ increases too. Because of the relation
$\Omega_{\chi} h^2 \propto <\sigma_{\mbi{ann}} v>^{-1}_{\mbi{int}}$
it follows that $R_{0,i}$ is large for neutralino configurations with
modest values of the relic abundance, and {\em vice versa}.

\subsection{Results for detection rates}

The most significant quantity in comparing experimental data and
theoretical evaluations for direct detection is the differential rate
$dR/dE_{ee} = (dR/dE_R)/Q$ (with $dR/dE_R$ defined in Eq.(\ref{eq:dir1}))
rather than the total rates, obtained by integration over wide ranges
of $E_{ee}$. By using the differential rate
instead of the integrated ones, one obtains the best signal--to--background
ratio. Note that the experimental spectra, apart from
an energy interval around  threshold, usually show a very flat
behaviour, whereas signals for light neutralinos are decreasing
functions of the nuclear recoil energy.

A complete procedure would then be to compare the
experimental and theoretical rates over the whole  $E_{ee}$ range.
However, to simplify the presentation here,
we give our results in terms of the rate integrated over a narrow
range  of 1 KeV at a specific value of $E_{ee}$, the one
which appears the most appropriate for each experiment:
typically it corresponds to a point close to the experimental
threshold. To be definite we consider the following cases:

i) Ge (natural composition). Among the various running experiments
\cite{mosca}, we select the
{two} which, at present, appear to provide the most stringent limits:
a) Caltech--PSI--Neuchatel \cite{[14]} with $E^{th}_{ee} = 2$ KeV,
differential rate $\simeq 3$ events/(Kg day KeV);
b) Heidelberg--Moscow \cite{klapdor} with $E^{th}_{ee} = 12$ KeV,
differential rate $ \simeq 0.2$ events/(Kg day KeV). Correspondingly,
for Ge we
have evaluated our rate by integrating $dR/dE_{ee}$ over the range
(2--3) KeV for experiment a) and over
(12--13)  KeV for experiment b). It turns out that the case b) provides
the most stringent bound also for light neutralinos.

ii)$^{129}$Xe. In this case, taking  into account the features of
the DAMA experiment \cite{[17]}, we have considered  the rate R
integrated over the range (40--41) KeV.

Our results are shown in Figs.16--19. Figs.16--18 report the rate
for a Ge detector for the regions of the parameter space which are
depicted in Figs.12--14, respectively. In parts (a) and (b) of each
figure, R is displayed in the form of a scatter plot, in terms of
$m_{\chi}$ and of the relic abundance, respectively. The
horizontal line denotes the present level of sensitivity
in the Heidelberg--Moscow experiment. We notice that, in all
cases shown in these figures, the experimental sensitivity is
already, for some configurations, at the level of the predicted rate.
Some points of the supersymmetric parameter space, denoted by filled squares in
Figs.12--14, are even already excluded by present data.
The exploration potential
of this class of experiments as the sensitivity is improved is apparent from
these figures.
Fig.19 shows the rate R for $^{129}$Xe for the region of the parameter
space displayed in Fig.13: again the horizontal line gives the
present experimental sensitivity. A comparison of Fig.19 with
Fig.17 shows that the Ge experiments are currently more effective.
However, it has to be noticed that experiments with
liquid Xe may become extremely competitive in the future \cite{[17]}.

A few more remarks are in order here:

i) The cases displayed in Figs.16--18 present the common feature of
providing fair
chances for direct detection. This is not a surprise,
since these representative points all belong to the category of
configurations with small values of $M_A$. As was stressed before,
once we move away from these appealing physical regions of the
neutralino parameter space, the rates for direct detection may fall
far below (by many orders of magnitude) the detection sensitivities
(present or future). This unfortunate situation occurs, for instance,
typically as we move towards smaller values of $\tb$. However, one
should keep in mind that the regime of very large $\tb$, where signals may be
sizeable, represents a very interesting scenario,
deserving much attention and exploration. In fact this is one of the two
options, very small or very large $\tb$, which seem to fit
low--energy phenomenology at the best \cite{ch}.

ii) The scatter plots in parts b) of Figs.16--19 show
explicitly  a property previously mentioned in Sect.IX.C,
namely that the scaling procedure adopted
to evaluate the neutralino local density implies  a
R--$\Omega_{\chi} h^2$ correlation.
Configurations which provide a measurable R usually entail a low
$\Omega$ and viceversa. Only in a few cases the neutralino may be detectable
by direct detection and also provide a sizeable contribution to
$\Omega$.

\section{Conclusions}
In the present paper we have discussed some possible scenarios for
neutralino dark matter which originate from the relaxation of the
assumption of strict universality for soft scalar masses at $M_{GUT}$.

This approach derives from the general consideration that many crucial
theoretical points entering not only grand unified and supersymmetric
theories, not to mention the Standard Model, are far from being understood
and/or verified. For this reason, any new theoretical assumption
has to be fully
scrutinized. This is even more important because new assumptions in
supersymmetric models are often introduced not because of solid
arguments, but rather for the sake of simplicity and
for the need to reduce  the large number of free parameters that
would otherwise prevent any firm prediction.

In our work we have  discussed different scenarios, by considering
various physical constraints in a sort of
hierarchical order, giving top priority to the requirement of
radiative EWSB, implemented with a no--fine--tuning criterion, and to the
cosmological relic neutralino density constraint.
Some other assumptions, often introduced in the
literature, have been
relaxed in our work. This is in particular the case for universality
in the soft scalar masses. However, it has to be remarked that the type of
departure from  universality that we have considered in our paper
is far from being
the most general one, as was noticed in Sect.II. In particular, it only
refers to the Higgs masses, and not to the sfermion masses.

The implications of the various scenarios on neutralino relic abundances
and rates for detection rates have been analysed, and the impact of a
non--universality in $m_0$ has been discussed for the whole range of
$\tb$. We have shown that the departure from $m_0$ universality is
particularly interesting in two respects:

i) Small values of
$M_A$ are allowed: this has in itself the dramatic consequence for
direct detection of generating a large value for the angle $\alpha$ and
large couplings to matter of the lightest neutralino $\chi$.

ii) Higgsino or mixed higgsino--gaugino configurations appear for all
$\tb$: this contrasts with the pure gaugino configurations
favoured by strict $m_0$ universality. \hfill \break
Consequences of such a departure from universality on the size of the
neutralino relic abundance have been
analysed for both large and small values of $\tan \beta$. It has been
shown that, because of the previous properties,
deviations from universality may reduce the value of
$\Omega_{\chi} h^2$.

The predicted rates for direct detection has been
analysed in detail and compared with current and foreseen experimental
sensitivities. The role of the previous properties in opening
interesting perspectives for this kind of search has been elucidated.
We find that presently--running experiments are already impacting interesting
regions of the neutralino parameters space in some of the non--universal
scenarios discussed here.

{\bf Acknowledgements.}
We wish to express our thanks to Uri Sarid for interesting discussions
and for contributions in the very early stages of our work. We also
gratefully acknowledge very useful conversations with Marek Olechowski
and informative discussions on experimental aspects of direct detection
with Pierluigi Belli, Rita Bernabei and Luigi Mosca. NF wishes to
express his gratitude to the Fondazione A. Della Riccia for a
fellowship. This work was supported in part by the Ministero della
Ricerca Scientifica e Tecnologica (Italy).

\vfill
\eject
{\bf Figure Captions}
\vspace{15 mm}

{\bf Figure 1} -- The ($m_{1/2}$, $\mu$) plane for $\tb =8$. The lines of
constant $m_\chi=$ 30 GeV, 60 GeV, 90 GeV are displayed as dashed lines. The
lines of constant P = 0.1, 0.5, 0.9, 0.99 are shown as solid lines. The dotted
region denotes the domain forbidden by present LEP data.
\vspace{10 mm}

{\bf Figure 2} -- Coefficients of the polynomial expressions (9--11)
as functions of $\tan \beta$: \hfill \break
a) coefficients of $M_{H_1}^2$: $a_1$ solid line, $b_1$ dashed line, $c_1$
dot--dashed line and $d_1$ dotted line, \hfill \break
b) coefficients of $M_{H_2}^2$: $a_2$ solid line, $b_2$ dashed line, $c_2$
dot--dashed line and $d_2$ dotted line, \hfill \break
c) coefficients of $\mu^2$: $J_1$ solid line, $J_2$ dashed line, $J_3$
dot--dashed line and $J_4$ dotted line, \hfill \break
d) coefficients of $M_A^2$: $K_1$ solid line, $K_2$ dashed line, $K_3$
dot--dashed line and $K_4$ dotted line.
\vspace{10 mm}

{\bf Figure 3} --
In the ($\delta_2$, $\delta_1$) plane, the lines where $J_2=0$ at fixed $\tan
\beta$ are displayed: $\tan\beta=53$, solid line;
$\tan\beta = 8$, dashed line; $\tan\beta=3$, dot--dashed line;
$\tan\beta = 2$, dotted line.
\vspace{10 mm}

{\bf Figure 4} --
In the ($\delta_2$, $\delta_1$) plane, the lines where $K_2=0$ at fixed $\tan
\beta$ are displayed: $\tan\beta=53$, solid line;
$\tan\beta = 40$, dashed line; $\tan\beta=8$, dot--dashed line.
\vspace{10 mm}

{\bf Figure 5} --
Graph of $M_A$ as a function of $\tan \beta$ at the representative point
$m_0 = 50$ GeV, $m_{1/2} = 200$ GeV.
\vspace{10 mm}

{\bf Figure 6} --
Scatter plot representing $\Omega_\chi h^2$ as a function of $m_\chi$ for $\tan
\beta = 8$, $\delta_1 = 0$ and $\delta_2 = 0$.
Parameters are varied on a linear equally--spaced
grid over the ranges: $10~\mb{GeV} \leq m_0 \leq 2~\mb{TeV}$,
$45~\mb{GeV} \leq m_{1/2} \leq 500~\mb{GeV}$.
\vspace{10 mm}

{\bf Figure 7} --
The same as in Figure 6, but with  $\delta_1 = -0.2$ and $\delta_2 = 0.4$.
\vspace{10 mm}

{\bf Figure 8} --
Scatter plot representing $\Omega_\chi h^2$ as a function of $m_\chi$ for $\tan
\beta = 53$, $\delta_1 = 0$ and $\delta_2 = 0$.
Parameters ranges are as in Figure 6.
\vspace{10 mm}

{\bf Figure 9} --
a) Parameter space in
($m_{1/2}$, $m_0$) plane for $\tan\beta=8$, $\delta_1=0$ and $\delta_2=0$.
Empty regions are excluded by:
i) accelerator constraints, ii) radiative EWSB conditions,
iii) neutralino is not the LSP.
Dots represent the region where $\Omega_\chi h^2 > 1$.
Regions with crosses are excluded by $b \rightarrow s\gamma$ and $m_b$
constraints.
In the regions denoted by squares, $0.01 <\Omega_\chi h^2 \leq 1$.
The region without fine--tuning is inside the box bounded by dashed lines.
\hfill \break
b) Parameter space represented in the ($m_{1/2}$, $\mu$) plane.
Solid lines correspond to the extreme values of $m_0$.
Notations are the same as in a), but crosses are omitted here.
\vspace{10 mm}

{\bf Figure 10} --
a) The same as in Figure 9, but with $\delta_1=-0.2$ and $\delta_2=0.4$.
In the regions denoted
by diamonds $\Omega_\chi h^2 \leq 0.01$. \hfill \break
b) Notations are the same as in a).
Note that the domain where the neutralino is the dark matter particle
with $0.01 < \Omega_\chi h^2 \leq 1$
has shifted to the higgsino--dominated region. Crosses are omitted here.
\vspace{10 mm}

{\bf Figure 11} --
The same as in Figure 9, but with $\delta_1=-0.8$ and $\delta_2=0.2$.
\vspace{10 mm}

{\bf Figure 12} --
a) Parameter space in the
($m_{1/2}$, $m_0$) plane for $\tan\beta=53$, $\delta_1=0$ and $\delta_2=0$.
Filled squares denote configurations excluded by direct detection with a Ge
detector \cite{klapdor}.
Other notations are as in Figure 9. \hfill \break
b) Parameter space represented in the ($m_{1/2}$, $\mu$) plane.
Notations are the same as in a).
\vspace{10 mm}

{\bf Figure 13} --
The same as in Figure 12, but with $\delta_1=0$ and $\delta_2=-0.2$.
\vspace{10 mm}

{\bf Figure 14} --
The same as in Figure 12, but with $\delta_1=0.7$ and $\delta_2=0.4$.
\vspace{10 mm}

{\bf Figure 15} --
a) Graphs of $M_A$ and $M_h$ as functions of $\tan \beta$. \hfill \break
b) Mixing angle $\alpha$ as a function of $\tan \beta$. \hfill \break
Here the  representative point is $m_0 = 100$ GeV, $m_{1/2} = 200$ GeV.
\vspace{10 mm}

{\bf Figure 16} --
Scatter plot of the rate for direct detection with a $Ge$ \cite{klapdor}
detector for $\tan\beta =53$, $\delta_1 = 0$ and $\delta_2 = 0$,
as a function of $m_\chi$ (a) and as a function of $\Omega_\chi h^2$ (b).
Parameters are varied on a linear equally--spaced
grid over the ranges: $10~\mb{GeV} \leq m_0 \leq 2~\mb{TeV}$,
$45~\mb{GeV} \leq m_{1/2} \leq 500~\mb{GeV}$.
\vspace{10 mm}

{\bf Figure 17} --
The same as in Figure 16, but with $\delta_1 = 0$ and $\delta_2 = -0.2$.
\vspace{10 mm}

{\bf Figure 18} --
The same as in Figure 16, but with $\delta_1 = 0.7$ and $\delta_2 = 0.4$.
\vspace{10 mm}

{\bf Figure 19} --
Scatter plot of the rate for direct detection with a $Xe$ \cite{[17]}
detector for $\tan\beta =53$, $\delta_1 = 0$ and $\delta_2 = -0.2$,
as a function of $m_\chi$ (a) and as a function of $\Omega_\chi h^2$ (b).
Parameters are varied on a linear equally--spaced
grid over the ranges: $10~\mb{GeV} \leq m_0 \leq 2~\mb{TeV}$,
$45~\mb{GeV} \leq m_{1/2} \leq 500~\mb{GeV}$.
\vspace{10 mm}


\begin{thebibliography}{99}
\bibitem{comm}
Neutralino dark matter has been analysed in the supergravity framework with
assumptions of exact unification at
$M_{GUT}$ in a number of papers.
In particular, in Ref.\cite{diehl}
the neutralino relic abundance and various detection rates have been analysed
in
this scheme using plausible but restrictive assumptions that represent one of
many possible physical scenarios.

\vspace{10 mm}
\end{thebibliography}
\end{document}